\documentclass{llncs}

\usepackage{graphicx}
\newenvironment{Nproof}{\vspace{-.5cm}\begin{trivlist}
   \item[] {\bf Proof:}}{\hspace{1.5em}\qed\end{trivlist}}

\newtheorem{fact}{Fact}

\newcommand{\eqref}[1]{{\rm(\ref{#1})}}

\newenvironment{smallenumerate}%
   {\vspace*{-4pt}
    \begin{enumerate}\itemsep=0pt}%
   {\end{enumerate}
    \vspace*{-2pt}}

\newenvironment{smallitemize}%
   {\vspace*{-4pt}
    \begin{itemize}\itemsep=0pt}%
   {\end{itemize}
    \vspace*{-2pt}}

\newcommand{\alg}[1]{\bigbreak\noindent{\bf #1}}
\newcommand{\algskip}{\itemsep=-8pt\baselineskip=0pt}
\newcommand{\myspace}{\hspace*{0.5cm}}
\newcommand{\lzsette}{{\tt LZ77}}

\newcommand{\lz}{{\tt LZ77}}
\newcommand{\LZZ}{{\tt LZ}}
\newcommand{\lzotto}{{\tt LZ78}}

\newcommand{\gzip}{{\tt gzip}}
\newcommand{\opt}{{\tt OPT}}
\newcommand{\ropt}{{\tt rOPT}}

\newcommand{\greedy}{{\cal G}}
\newcommand{\prun}{\widetilde{\cal G}}

\newcommand{\Seq}{{\cal P}}
\newcommand{\lcp}{{\tt lcp}}
\newcommand{\lca}{{\tt lca}}

\newcommand{\DAG}{{\tt DAG}}
\newcommand{\SSSP}{{\tt SSSP}}
\newcommand{\T}{{\cal T}}
\newcommand{\W}{{\cal W}}
\newcommand{\nil}{{\tt nil}}
\newcommand{\p}{{\tt parent}}
\newcommand{\mm}{{\tt min}}
\newcommand{\MM}{{\tt max}}
\newcommand{\maxpos}{{\tt mp}}

\title{Bit-Optimal Lempel-Ziv compression\thanks{It has been partially supported by {\sf
Yahoo! Research}, Italian MIUR Italy-Israel FIRB Project and PRIN
MAINSTREAM. The authors' address is Dipartimento di Informatica,
L.go B. Pontecorvo 3, 56127 Pisa, Italy. email:
\{ferragina,nitto,rossano\}@di.unipi.it}}

\author{Paolo Ferragina\inst{1} \and Igor Nitto\inst{1} \and Rossano Venturini\inst{1}}

\institute{Dipartimento di Informatica, Universit\`a di Pisa, Italy}

\begin{document}

\maketitle

\begin{abstract}
One of the most famous and investigated lossless data-compression
scheme is the one introduced by Lempel and Ziv about 40 years ago
\cite{lz77}. This compression scheme is known as "dictionary-based
compression" and consists of squeezing an input string by replacing
some of its substrings with (shorter) codewords which are actually
pointers to a dictionary of phrases built as the string is
processed. Surprisingly enough, although many fundamental results
are nowadays known about upper bounds on the speed and effectiveness
of this compression process (see e.g. \cite{koma00,RS:02} and
references therein), {\em ``we are not aware of any parsing scheme
that achieves optimality when the \lzsette-dictionary is in use
under any constraint on the codewords other than being of equal
length''} \cite[pag. 159]{RS:02}. Here optimality means to achieve
the {\em minimum} number of bits in compressing {\em each
individual} input string, {\em without} any assumption on its
generating source. In this paper we provide the first LZ-based
compressor which computes the bit-optimal parsing of {\em any} input
string in efficient time and optimal space, for a general class of
variable-length codeword encodings which encompasses most of the
ones typically used in data compression and in the design of search
engines and compressed indexes
\cite{NM07,Salomon04,Witten:1999:MGC}.
\end{abstract}

\section{Introduction} \label{sec:Intro}
 The problem of {\em lossless} data compression
consists of compactly representing data in a format that can be
faithfully recovered from the compressed file. Lossless compression
is achieved by taking advantage of the {\em redundancy} which is
often present in the data generated by either humans or machines.
One of the most famous lossless data-compression scheme is the one
introduced by Lempel and Ziv in the late 70s~\cite{lz77,lz78}. It
has been ``the solution'' to lossless compression for nearly 15
years, and indeed many (non-)commercial programs are currently based
on it-- like {\tt gzip}, {\tt zip}, {\tt pkzip}, {\tt arj}, {\tt
rar}, just to cite a few. This compression scheme is known as {\em
dictionary-based compression}, and consists of squeezing an input
string $S[1,n]$ by replacing some of its substrings (phrases) with
(shorter) {\em codewords} which are actually pointers to a
dictionary being either {\em static} (in that it has been
constructed before the compression starts) or {\em dynamic} (in that
it is built as the input string is compressed). The well-known
\lzsette\ and \lzotto\ compressors, proposed by Lempel and Ziv in
\cite{lz77,lz78}, and all their numerous variants \cite{Salomon04},
are interesting examples of {\em dynamic} dictionary-based
compression algorithms. In \lzsette, and its variants, the
dictionary consists of all substrings occurring in the previously
scanned portion of the input string, and each codeword consists of a
triple $\langle d,\ell,c\rangle$ where $d$ is the relative {\em
offset} of the copied phrase, $\ell$ is its length and $c$ is the
single (new) character following it. In \lzotto, the dictionary is
built upon phrases extracted from the previously scanned prefix of
the input string, and each codeword consists of a pair $\langle {\tt
id}, c\rangle$ where {\tt id} is the {\em identifier} of the copied
phrase in the dictionary and $c$ is the character following that
phrase in the subsequent suffix of the string.

Many theoretical and experimental results have been dedicated to
LZ-compressors  in these thirty years (see e.g.~\cite{Salomon04} and
references therein); and, although today there are alternative
solutions to the problem of lossless data compression potentially
offering better compression bounds (e.g., Burrows-Wheeler
compression and Prediction by Partial Matching
\cite{Witten:1999:MGC}), dictionary-based compression is still
widely used in everyday applications because of its unique
combination of compression power and speed. Over the years
dictionary-based compression has also gained importance as a general
algorithmic tool, being employed in the design of compressed
text indexes \cite{NM07}, or in {\em universal} clustering tools
\cite{Vitanyi03}, or in designing {\em optimal} pre-fetching
mechanisms \cite{VitterK96}.

Surprisingly enough some important problems on the combinatorial
properties of the {\em LZ-parsing} are still open, and they will be
the main topic of investigation of this paper. Take the classical
\lzsette\ and \lzotto\ algorithms. They adopt a {\em greedy parsing}
of the input string: namely, at each step, they take the {\em
longest} dictionary phrase which is a prefix of the currently
unparsed string suffix. It is well-known \cite{RS:02} that greedy
parsing is optimal with respect to the {\em number of phrases} in
which $S$ can be parsed by any suffix-complete dictionary (like
\lzsette); and a small variation of it (called {\em
flexible}-parsing \cite{FP99}) is optimal for prefix-complete
dictionaries (like \lzotto). Of course, the number of parsed phrases
influences the compression ratio and, indeed, various authors
\cite{lz77,lz78,koma00} proved that greedy parsing achieves
asymptotically the {\em (empirical) entropy} of the source
generating the input string $S$. However, these fundamental results
have {\em not yet closed} the problem of optimally compressing $S$
because the optimality in the number of parsed phrases {\em is not
necessarily equal} to the optimality in the number of bits output by
the final compressor. Clearly, if the phrases are compressed via an
{\em equal-length} encoder, like in \cite{Salomon04,lz77,koma00},
then the produced output is {\em bit optimal}. But if one aims for
higher compression by using {\em variable-length encoders} for the
parsed phrases (see e.g. \cite{Witten:1999:MGC,univ}), then the
bit-length of the compressed output produced by the greedy-parsing
scheme is {\em not necessarily optimal}.

As an illustrative example, consider the \lzsette-compressor and
assume that the copy of the $i$th phrase occurs very far from the
current unparsed position, and thus its $d$-value is large. In this
case it could be probably more convenient to renounce to the {\em
maximality} of that phrase and split it into (several) smaller
phrases which possibly occur closer to that position and thus can be
encoded in fewer bits overall. Several solutions are indeed known
for determining the bit-optimal parsing of $S$, but they are either
inefficient \cite{Schuegraf,Mignosi07} taking $\Theta(n^2)$ time and
space in the worst case, or approximate \cite{Raita89}, or they rely
on heuristics \cite{klein,heur1,heur3,heur4,Mignosi07} which do not
provide any guarantee on the time/space performance of the
compression process. This is the reason why Rajpoot and Sahinalp
stated in \cite[pag. 159]{RS:02} that ``We are not aware of any
on-line or off-line parsing scheme that achieves optimality when the
\lzsette-dictionary is in use under any constraint on the codewords
other than being of equal length''.

Motivated by these poor results, we address in this paper the
question posed by Rajpoot and Sahinalp by investigating a general
class of variable-length codeword encodings which are typically used
in data compression and in the design of search engines and
compressed indexes \cite{NM07,Salomon04,Witten:1999:MGC}. We prove
that the classic greedy-parsing scheme deploying these encoders may
be far from the bit-optimal parsing by a multiplicative factor
$\Omega(\log n/\log \log n)$, which is indeed unbounded
asymptotically (Section \ref{sec:LZopt}, Lemma \ref{lem:example}).
This result is obtained by considering an infinite family of strings
$S$ of increasing length $n$ and low empirical entropy, and by showing
that for these strings copying the longest phrase, as \lzsette\ does,
may be dramatically inefficient. We notice that this gap between
\lzsette\ and the bit-optimal compressor strengthen the results proved
by Kosaraju and Manzini in \cite{koma00}, who showed
that the compression rate of \lzsette\ converges asymptotically to the $k$th order empirical
entropy $H_k(S)$ of string $S$, and this rate is dominated for low entropy
strings by an additive term $O(\frac{\log\log n}{\log n})$
which depends on the string length $n$ and not on its compressibility
$H_k(S)$. These are properly the strings for which the bit-optimal parser
is much better than \lzsette, and thus closer to the entropy of $S$.

Given these premises, we investigate and design an LZ-based compressor that computes the bit-optimal
parsing for those variable-length integer encoders in efficient time and optimal space, in
the worst case (Section \ref{sec:efficient}, Theorem
\ref{teo:general}). Due to space limitations, we will detail our
results only for the \lzsette-dictionary, and defer the discussion
on other dictionary-based schemes (like \lzotto) to the last Section
\ref{sec:conclusion}.
Technically speaking, we follow \cite{Schuegraf} and model the
search for a bit-optimal parsing of an input string $S[1,n]$, as a
{\em single-source shortest path} problem  (shortly, \SSSP) on a
{\em weighted {\tt DAG}} $\greedy(S)$ consisting of $n$ nodes, one
per character of $T$, and $e$ edges, one per possible
\lzsette-parsing step. Every edge is weighted according to the
length in bits of the codeword adopted to compress the corresponding
\lzsette-phrase. Since LZ-codewords are tuples of integers (see
above), we consider in this paper a class of codeword encoders which
satisfy the so called {\em increasing cost property}: the larger is
the integer to be encoded, the longer is the codeword. This class
encompasses most of the encoders frequently used in the literature
to design data compressors \cite{univ}, compressed full-text indexes
\cite{NM07} and search engines \cite{Witten:1999:MGC}. We prove new
combinatorial properties for the \SSSP-problem formulated on the
graph $\greedy(S)$ weighted according to these encoding functions
and show that, unlike \cite{Schuegraf} (for which $e = \Theta(n^2)$
in the worst case), the computation of the \SSSP\ in $\greedy(S)$
can be restricted onto a subgraph $\prun(S)$ whose size is {\em
provably smaller} than the complete graph (see Theorem
\ref{teo:pruning}). Actually, we show that the size of $\prun(S)$ is
related to the {\em structural features} of the integer-encoding
functions adopted to compress the LZ-phrases (Lemma
\ref{lem:subgraph}). Finally, we design an algorithm that computes
the \SSSP\ of $\prun(S)$ without materializing that subgraph {\em
all at once}, but by creating and exploring its edges on-the-fly in
optimal $O(1)$ amortized time per edge and using $O(n)$ optimal
space overall. As a result, our \lzsette-compressor achieves optimal
compression ratio, by using optimal $O(n)$ working space and taking
time proportional to $|\prun(S)|$ (hence, it is optimal in its size).

If the {\tt LZ}-phrases are encoded with equal-length codewords, our
approach is optimal in compression ratio and time/space performance,
as it is classically known \cite{Salomon04}. But if we consider the
more general (and open) case of the variable-length Elias or
Fibonacci codes, as it is typical of data compressors and compressed
indexes \cite{univ,NM07}, then our approach is optimal in
compression ratio and working-space occupancy, and takes $O(n \log
n)$ time in the worst case. Most other variable-length integer
encoders fall in this case too (see e.g. \cite{Witten:1999:MGC}). To
the best of our knowledge, this is the first result providing a {\em positive answer} to Rajpoot-Sahinalp's
question above!

The final Section \ref{sec:conclusion} will discuss variations and
extensions of our approach, considering the cases of a {\em bounded}
compression-window and of other suffix- or prefix-complete
dictionary construction schemes (like \lzotto).

\section{On the Bit-Optimality of LZ-parsing}
\label{sec:LZopt}

Let $S[1,n]$ be a string drawn from an alphabet $\Sigma$ of size
$\sigma$. We will use $S[i]$ to denote the $i$th symbol of $S$;
$S[i:j]$ to denote the substring (also called the {\em phrase})
extending from the $i$th to the $j$th symbol in $S$ (extremes
included); and $S_i=S[i:n]$ to denote the $i$-th suffix of $S$.

Dictionary-based compression works in two intermingled phases: {\em
parsing} and {\em encoding}. Let $w_1, w_2, \ldots, w_{i-1}$ be the
phrases in which a prefix of $S$ has been already parsed. The parser
selects the next phrase $w_i$ as one of the phrases in the current
dictionary that prefix the remaining suffix of $S$, {\em and}
possibly attaches to this phrase few other following symbols
(typically one) in $S$. This addition is sometimes needed to enrich
the dictionary with new symbols occurring in $S$ and never
encountered before in the parsing process. Phrase $w_i$ is then
represented via a proper {\em reference} to the {\em dynamic}
dictionary, which is built during the parsing process. The
well-known \lzsette\ and \lzotto\ compressors \cite{lz77,lz78}, and
all their variants \cite{Salomon04}, are incarnations of
the above compression scheme and differentiate themselves mainly by
the way the dynamic dictionary is built, by the rule adopted to
select the next phrase, and by the encoding of the dictionary
references.

In the rest of the paper we will concentrate on the \lzsette-scheme
and defer the discussion of \lzotto's, and other approaches, to the
last Section \ref{sec:conclusion}. In \lzsette\ and its variants,
the dictionary consists of all substrings starting in the scanned
prefix of $S$,\footnote{Notice that it admits the {\em overlapping}
between the current dictionary and the next phrase to be
constructed.} and thus it consists (implicitly) of a {\em quadratic}
number of phrases which are (explicitly) represented via a (dynamic)
indexing data structure that takes linear space in $|S|$ and
efficiently searches for repeated substrings. The
\lzsette-dictionary satisfies two main properties: {\em Prefix
Completeness}--- i.e. for any given dictionary phrase, all of its
prefixes are also dictionary phrases--- and {\em  Suffix
Completeness}--- i.e. for any given dictionary phrase, all of its
suffixes are also dictionary phrases. At any step, the
\lzsette-parser proceeds by adopting the so called {\em longest
match heuristic}: that is, $w_i$ is taken as the \textit{longest}
phrase of the current dictionary which prefixes the remaining suffix
of $S$. This will be hereafter called {\em greedy parsing}. The
classic \lzsette-parser finally adds one further symbol to $w_i$
(namely, the one following the phrase in $S$) to form the next
phrase of $S$'s parsing.

In the rest of our paper, and without loss of generality, we
consider the \lzsette-variant which avoids the additional symbol per
phrase, and thus represents the next phrase $w_i$ by the integer
pair $\langle d_i,\ell_i\rangle$ where $d_i$ is the relative {\em
offset} of the copied phrase $w_i$ within the prefix $w_1\cdots
w_{i-1}$ and $\ell_i$ is its length (i.e. $|w_i|$). We notice that
every first occurrence of a new symbol $c$ is encoded as $\langle
0,c\rangle$. Once phrases are identified and represented via pairs
of integers, their components are compressed via {\em
variable-length integer encoders} which will eventually produce the
compressed output of $S$ as a sequence of bits.

In order to study and design bit-optimal parsing schemes, we
therefore need to deal with integer encoders. Let $f$ be an
integer-encoding function that maps any integer $x \in [n]$ into a
(bit-)codeword $f(x)$ whose length (in bits) is denoted by $|f(x)|$.
In this paper we consider variable-length encodings which use longer
codewords for larger integers:

\begin{property}[Increasing Cost Property]
\label{prop:Incr} For any $x,y \in [n]$ it is $x \leq y$ iff
$|f(x)|\leq |f(y)|$.
\end{property}

This property is satisfied by most practical integer encoders
\cite{Witten:1999:MGC}, as well by equal-length codewords, Elias
codes (i.e. gamma, delta, and their derivatives \cite{elias}), and
Fibonacci's codes \cite{univ}. Therefore, this class encompasses all
encoders typically used in the literature to design data compressors
\cite{Salomon04}, compressed full-text indexes \cite{NM07} and
search engines \cite{Witten:1999:MGC}.

Given two integer-encoders $f$ and $g$ (possibly $f=g$) which
satisfy the Increasing Cost Property \ref{prop:Incr}, we denote by
$\LZZ_{f,g}(S)$ the compressed output produced by the greedy-parsing
strategy in which we have used $f$ to compress the distance $d_i$,
and $g$ to compress the length $\ell_i$ of any parsed phrase $w_i$.
$\LZZ_{f,g}(S)$ thus encodes any phrase $w_i$ in
$|f(d_i)|+|g(\ell_i)|$ bits. We have already noticed that
$\LZZ_{f,g}(S)$ is not necessarily bit optimal, so we will hereafter
use $\opt_{f,g}(S)$ to denote the {\em $(f,g)$-optimal parser},
namely the one that parses $S$ into a sequence of phrases which are
drawn from the \lzsette-dictionary and which {\em minimize} the
total number of bits produced by their encoders $f$ and $g$. Given
the above observations it is immediate to infer that
$|\LZZ_{f,g}(S)| \geq |\opt_{f,g}(S)|$. However this bound does not
provide us with any estimate of how much worse the greedy parsing
can be with respect to $\opt_{f,g}(S)$. In what follows we identify
an infinite family of strings for which the compressed output of the
greedy parser is a multiplicative factor $\Omega(\log n/\log \log
n)$ worse than the bit-optimal parser. This result shows that the
ratio $\frac{|\LZZ_{f,g}(S)|}{|\opt_{f,g}(S)|}$ is indeed
asymptotically unbounded, and thus poses the serious need for an
$(f,g)$-optimal parser, as clearly requested by \cite{RS:02}.

Our argument holds for any choice of $f$ and $g$ from the family of
encoding functions that represent an integer $x$ with a bit string
of size $\Theta(\log x)$ bits. (Thus the well-known Elias' and
Fibonacci's coders belong to this family.)
Taking inspiration from the proof of Lemma 4.2 in \cite{koma00},
we consider the infinite
family of strings $S_l=ba^l\;c^{2^{l}}\;ba\;ba^2\;ba^3 \ldots ba^l$,
parameterized in the positive value $l$. The \lz-parser partitions
$S_l$ as\footnote{Recall the variant of \lzsette\ we are considering
in this paper, which uses just a pair of integers per phrase, and
thus drops the char following that phrase in $S$.}

\[(b)\;(a)\;(a^{l-1})\;(c)\;(c^{2^l-1})\;(ba)\;(ba^2)\;(ba^3)\; \ldots\; (ba^l)\]
where the symbols forming a parsed phrase have been delimited within
a pair of brackets. \lz\ thus copies the latest $l$ phrases from the
beginning of $S_l$ and takes at least $l\: |f(2^l)| = \Theta(l^2)$
bits. Let us now consider a more parsimonious parser, called \ropt,
which selects the copy of $ba^{i-1}$ (with $i > 1$) from its
immediately previous occurrence:

\[(b)\;(a)\;(a^{l-1})\;(c)\;(c^{2^l-1})\;(b)\;(a)\;(ba)\;(a)\;(ba^2)\;(a)\;
\ldots\; (ba^{l-1})\;(a) \] $\ropt(S_l)$ takes $|g(2^l)|+|g(l)|
+ \sum_{i=2}^l [|f(i)|+|g(i)|+f(0)]+O(l) = O(l\log l)$ bits. Since
$\opt(S_l) \leq \ropt(S_l)$, we can conclude that

\begin{equation}\label{eqn:example}
\frac{|\LZZ_{f,g}(S_l)|}{|\opt_{f,g}(S_l)|} \geq
\frac{|\LZZ_{f,g}(S_l)|}{|\ropt(S_l)|} \geq \Theta\left(\frac{l}{\log l}\right)
\end{equation}

Since $|S_l| = 2^l + l^2 - O(l)$, we have that $l =\Theta(\log |S_l|)$
for sufficiently long strings. Using this estimate into
Inequality \ref{eqn:example} we finally obtain:

\begin{lemma}
\label{lem:example} There exists an infinite family of strings such
that, for any of its elements $S$, it is $|\LZZ_{f,g}(S)| \geq
\Theta(\log |S|/\log \log |S|) \: |\opt_{f,g}(S)|$.
\end{lemma}

On the other hand, we can prove that this lower bound is tight up to
a $\log \log |S|$ multiplicative factor, by easily extending to
\lzsette-dictionary and Property \ref{prop:Incr}, a result proved in
\cite{Raita92} for static dictionaries. Precisely, we can show that
$\frac{|\LZZ_{f,g}(S)|}{|\opt_{f,g}(S)|} \leq
\frac{|f(n)|+|g(n)|}{|f(0)|+|g(0)|}$, which is upper bounded by
$O(\log n)$ because $|S|=n$, $|f(n)|=|g(n)|=\Theta(\log n)$ and
$|f(0)|=|g(0)|=O(1)$.

\section{Bit-Optimal Parsing and \SSSP-problem}
\label{sec:opt}

Following \cite{Schuegraf}, we model the design of a bit-optimal
\lzsette-parsing strategy for a string $S$ as a Single-Source
Shortest Path problem (shortly, \SSSP-problem) on a weighted {\tt
DAG} $\greedy(S)$ defined as follows. Graph $\greedy(S)=(V,E)$ has
one vertex per symbol of $S$ plus a dummy vertex $v_{n+1}$, and its
edge set $E$ is defined so that $(v_i,v_j)\in E$ iff (1) $j = i+1$
or (2) the substring $S[i:j-1]$ occurs in $S$ starting from a
(previous) position $p < i$ (clearly $i < j$ and thus $\greedy(S)$
is a DAG). Every edge $(v_i,v_j)$ is labeled with the pair $\langle
d_{i,j}, \ell_{i,j} \rangle$ where

\begin{smallitemize}
\item $d_{i,j} = i-p$ is the distance between $S[i:j-1]$ and the position $p$ of its
right-most copy in $S$. We set $d_{i,j}=0$ whenever $j=i+1$.

\item We set $\ell_{i,j} = j-i$, if $j > i+1$, or $\ell_{i,j} = S[i]$ otherwise.
\end{smallitemize}

It is easy to see that the edges outgoing from $v_i$ denote all
possible parsing steps that can be taken by any parsing strategy
which uses a \lzsette-dictionary. Hence, there exists a one-to-one
correspondence between paths from $v_1$ to $v_{n+1}$ in $\greedy(S)$
and parsings of the whole string $S$: any path $\pi = v_{1}\cdots
v_{n+1}$ corresponds to the parsing $\Seq_\pi(S)$ represented by the
phrases labeling the edges traversed by $\pi$. If we weight every
edge $(v_i,v_j) \in E$ with an integer $c(v_i,v_j) = |f(d_{i,j})| +
|g(\ell_{i,j})|$ which accounts for the cost of encoding its label
(phrase) via the encoding functions $f$ and $g$, then the length in
bits of the encoded parsing $\Seq_\pi(S)$ is equal to the cost of
the weighted path $\pi$ in $\greedy(S)$. We have therefore reduced
the problem of determining $\opt_{f,g}(S)$ to the problem of
computing the \SSSP\ of $\greedy(S)$ from $v_1$ to $v_{n+1}$.

Given that $\greedy(S)$ is a \DAG, its shortest path from $v_1$ to
$v_{n+1}$ can be computed in $O(|E|)$ time and space. In the worst
case (take e.g. $S=a^n$), this is $\Theta(n^2)$, and thus it is
inefficient and unusable in practice \cite{Schuegraf,Raita89}. In
what follows we show that the computation of the \SSSP\ can be
actually restricted to a subgraph of $\greedy(S)$ whose size is
provably $o(n^2)$ in the worst case, and typically $O(n \log n)$ for
most known integer-encoding functions. Then we will design efficient
and sophisticated algorithms and data structures that will allow us
to generate this subgraph {\em on-the-fly} by taking $O(1)$
amortized time per edge and $O(n)$ space overall. These algorithms
will be therefore time-and-space optimal for the subgraph in hand!

\section{An useful, small, subgraph of $\greedy(S)$}
\label{sub:subgraph}

We use $FS(v)$ to denote the {\em forward star} of a vertex $v$,
namely the set of vertices pointed to by $v$ in $\greedy(S)$; and we
use $BS(v)$ to denote the {\em backward star} of $v$, namely the set
of vertices pointing to $v$  in $\greedy(S)$. By construction of
$\greedy(S)$, for any $v_j \in FS(v_i)$ it is $i<j$; so that, all of
the edges are oriented rightward, and in fact $\greedy(S)$ is a DAG.
We can actually show a stronger property on the distribution of the
indices of the vertices in $FS(v)$ and $BS(v)$, namely that they
form a {\em contiguous} range.

\begin{fact}\label{fact:contiguous}
Given a vertex $v_i$, it is $FS(v_i)=\{v_{i+1} \ldots, v_{i+x-1}, v_{i+x}\}$ and $BS(v_i)=\{v_{i-y} \ldots, v_{i-2}, v_{i-1}\}$. Note that $x,y$ are smaller than the length of the longest
repeated substring in $S$.
\end{fact}
\begin{Nproof}
By definition of $(v_i,v_{i+x})$, string $S[i:i+x-1]$ occurs at some
position $p < i$ in $S$. Any prefix $S[i:k-1]$ of $S[i:i+x-1]$ also
occurs at that position $p$, thus $v_k \in FS(v_i)$. The bound on
$x$ immediately derives from the definition of $(v_i,v_{i+x})$. A
similar argument derives the property on $BS(v_i)$.
\end{Nproof}

This actually means that if an edge does exist in $\greedy(S)$, then
they do exist also all edges which are {\em nested} within it and
are incident into one of its extremes. The following property
relates the indices of the vertices $v_j \in FS(v_i)$ with the cost
of their connecting edge $(v_i,v_j)$, and not surprisingly shows
that the smaller is $j$ (i.e. shorter edge), the smaller is the cost
of encoding the parsed phrase $S[i:j-1]$.\footnote{Recall that
$c(v_i,v_j) = |f(d_{i,j})| + |g(\ell_{i,j})|$, if the edge does
exist, otherwise we set $c(v_i,v_j) = +\infty$.}

\begin{fact} \label{fact:FS}
Given a vertex $v_i$, for any pair of vertices $v_{j'},v_{j''}\in
FS(v_i)$ such that $j' < j''$, we have that $c(v_i, v_{j'}) \leq
c(v_i,v_{j''})$. The same property holds for $v_{j'},v_{j''}\in
BS(v_i)$.
\end{fact}
\begin{Nproof}
We have that $d_{i,j'} \leq d_{i, j''}$ and $\ell_{i,j'} <
\ell_{i,j''}$ because $S[i: j'-1]$ is a prefix of $S[i:j''-1]$ and
thus the first substring occurs wherever the latter occurs. The
property holds because $f$ and $g$ satisfy the Increasing Cost
Property \ref{prop:Incr}.
\end{Nproof}

Given these monotonicity properties, we are ready to characterize a
special subset of the vertices in $FS(v_i)$, and their connecting
edges.

\begin{definition}\label{def:maximal}
An edge $(v_i,v_j)\in E$ is called
\begin{smallitemize}
\item {\em $d-$maximal} iff the next edge from $v_i$ takes more bits to encode its distance. Namely,
we have that $|f(d_{i,j})| < |f(d_{i,j+1})|$.
\item {\em $\ell-$maximal} iff the next edge from $v_i$ takes more bits to encode its length. Namely,
we have that $|g(l_{i,j})| < |g(l_{i,j+1})|$.
\end{smallitemize}
\noindent Overall, we say that edge $(v_i,v_j)$ is {\em maximal} if
it is either $d$-maximal or $\ell$-maximal: thus $c(v_i,v_{j}) <
c(v_i,v_{j+1})$.
\end{definition}

Now, we wish to count the number of maximal edges outgoing from
$v_i$. This number clearly depends on the integer-encoding functions
$f$ and $g$ (which satisfy Property \ref{prop:Incr}). Let $Q(f,n)$
(resp. $Q(g,n)$) be the number of different codeword lengths
generated by $f$ (resp. $g$) when applied to integers in the range
$[n]$. We can partition $[n]$ in contiguous sub-ranges $I_1, I_2,
\ldots, I_{Q(f,n)}$ such that the integers in $I_i$ are mapped by
$f$ to codewords {\em (strictly) shorter than} the codewords for the
integers in $I_{i+1}$. Similarly, $g$ partitions the range $[n]$ in
$Q(g,n)$ contiguous sub-ranges.

\begin{lemma} \label{lem:Qedges}
There are at most $Q(f,n) + Q(g,n)$ maximal
edges outgoing from any vertex $v_i$.
\end{lemma}
\begin{Nproof}
By Fact \ref{fact:contiguous}, vertices in $FS(v_i)$
have indices in a range $R$, and by Fact \ref{fact:FS},
$c(v_i,v_j)$ is monotonically non-decreasing as $j$ increases in
$R$. Moreover we know that $f$ (resp. $g$) cannot change more than
$Q(f,n)$ (resp. $Q(g,n)$) times, so that the statement follows.
\end{Nproof}

In order to speed up the computation of a \SSSP\ connecting $v_1$ to
$v_{n+1}$ in $\greedy(S)$, we construct a subgraph $\prun(S)$ which is
provably smaller than $\greedy(S)$ and contains one of those \SSSP. Next theorem shows that
$\prun(S)$ can be formed by taking just the maximal edges of
$\greedy(S)$.

\begin{theorem}\label{teo:pruning}
There exists a shortest path in $\greedy(S)$ connecting $v_1$ to
$v_{n+1}$ and traversing only maximal edges.
\end{theorem}
\begin{Nproof}
By contradiction assume that every such shortest paths contains at
least one non-maximal edge. Let $\pi = v_{i_1}v_{i_2} \ldots
v_{i_k}$, with $i_1 = 1$ and $i_k=n+1$, be one of these shortest
paths, and let $\gamma = v_{i_1}\ldots v_{i_r}$ be the longest
initial subpath of $\pi$ which traverses only maximal edges. Assume
w.l.o.g.\ that $\pi$ is the shortest path maximizing the value of
$|\gamma|$. We know that $(v_{i_r}, v_{i_{r+1}})$ is a non-maximal
edge, and thus we can take the maximal edge $(v_{i_r}, v_j)$ that
has the same cost. By definition of maximal edge, it is $j >
i_{r+1}$; furthermore, we must have $j<n+1$ because we assumed that
no path is formed only by maximal edges. Thus it must exist an index
$i_h \geq i_r$ such that $j \in [i_h,i_{h+1}]$, because indices in
$\pi$ are increasing given that $\greedy(S)$ is a DAG. Since
$(v_{i_h},v_{i_{h+1}})$ is an edge of $\pi$, by Fact
\ref{fact:contiguous} follows that it does exist the edge
$(v_j,v_{i_{h+1}})$, and by Fact \ref{fact:FS} on $BS(v_{i_{h+1}})$
we can conclude that $c(v_{j},v_{i_{h+1}}) \leq c(v_{i_{h}},
v_{i_{h+1}})$. Consequently, the path $v_{i_1}\cdots v_{i_r} v_j
v_{i_{h+1}} \cdots v_{i_k}$ is also a shortest path but its longest
initial subpath of maximal edges consists of $|\gamma|+1$ vertices,
which is a contradiction!
\end{Nproof}

Theorem \ref{teo:pruning} implies that the distance between $v_1$
and $v_{n+1}$ is the same in $\greedy(S)$ and $\prun(S)$, with the
advantage that computing distances in $\prun(S)$ can be done faster
and in reduced space, because of its smaller size. In fact, Lemma
\ref{lem:Qedges} implies that $|FS(v)| \leq Q(f,n) + Q(g,n)$, so
that

\begin{lemma}
\label{lem:subgraph} Subgraph $\prun(S)$ consists of $n+1$ vertices
and at most $n(Q(f,n) + Q(g,n))$ edges.
\end{lemma}

For Elias' codes \cite{elias}, Fibonacci's codes \cite{univ}, and
most practical integer encoders used for search engines and data
compressors \cite{Salomon04,Witten:1999:MGC}, it is $Q(f,n) = Q(g,n)
= O(\log n)$. Therefore $|\prun(S)| = O(n \log n)$ and hence it is
provably smaller than the complete graph built and used by the
previous papers \cite{Schuegraf,Raita89,Mignosi07}.

The next technical step consists of achieving time efficiency and
optimality in working space, because we cannot construct $\prun(S)$
all at once. In the next sections we design an algorithm that
generates $\prun(S)$ {\em on-the-fly} as the computation of its
\SSSP\ goes on, and pays $O(1)$ amortized time per edge and no more
than $O(n)$ optimal space overall. In some sense, this algorithm is
optimal for the identified sub-graph $\prun(S)$.

\section{A bit-optimal parser}
\label{sec:efficient}

\begin{figure}[t]
\hrule \small {\algskip \alg{Optimal-Parser($S[1,n]$)}
\begin{smallenumerate}
\item $C[1] = 0;$ $P[1] = 1;$
\item {\bf for each} $i \in [2,n+1]$ {\bf do} $C[i] = +\infty;$ $P[i] = NIL;$
\item {\bf for} $i= 1$ {\bf to} $n$ {\bf do}
\item \myspace {\bf generate} on-the-fly all maximal edges in $FS(v_i)$;
\item \myspace {\bf for any} $(v_i,v_j)$ maximal {\bf do}
\item \myspace \myspace {\bf if} $C[j] > C[i] + c(v_i,v_j)$ {\bf then} $C[j] = C[i] + c(v_i,v_j);$ $P[j] = i;$
\end{smallenumerate}
} \hrule\caption{\small Algorithm to compute the \SSSP\ of the
subgraph $\prun(S)$.\label{fig:algo-scan}}
\end{figure}

From a high level, our solution proceeds as in Figure
\ref{fig:algo-scan}, where a variant of a classic linear-time
algorithm for \SSSP\ over a \DAG\ is reported \cite[Section
24.2]{CLR}. In that pseudo-code entries $C[i]$ and $P[i]$ hold,
respectively, the shortest path distance from $v_1$ to $v_i$ and the
predecessor of $v_i$ in that shortest path. The main idea of {\tt
Optimal-Parser} consists of scanning the vertices of $\prun(S)$ in
topological order, and of generating {\em on-the-fly} and {\em
relaxing} (Step 6) the edges outgoing from a vertex $v_i$ only when
$v_i$ becomes the current vertex. The correctness of {\tt
Optimal-Parser} follows directly from Theorem 24.5 of \cite{CLR} and
our Theorem \ref{teo:pruning}.

The key difficulty in this process consists of how to {\em generate
on-the-fly and efficiently (in time and space) the maximal edges
outgoing from vertex $v_i$}. We will refer to this problem as the
{\em forward-star generation} problem, and use \texttt{FSG} for
brevity. In the next section we show that, when $\sigma \leq n$,
\texttt{FSG} takes $O(1)$ amortized time per edge and $O(n)$ space
in total. As a result (Theorem \ref{teo:general}),
\texttt{Optimal-Parser} requires $O(n(Q(f,n) + Q(g,n)))$ time in the
worst case, since the main loop is repeated $n$ times and we have no
more than $Q(f,n) + Q(g,n)$ maximal edges per vertex (Lemma
\ref{lem:Qedges}). The space used is that for the
\texttt{FSG}-computation plus the two arrays $C$ and $P$; hence, it
will be shown to be $O(n)$ in total. In case of a large alphabet
$\sigma> n$, we need to add $T_{sort}(n,\sigma)$ time because of
the sorting/remapping of $S$'s symbols into $[n]$.

\begin{theorem}\label{teo:general}
Given a string $S[1,n]$ drawn from an alphabet of size $\sigma$, and
two integer-encoding functions $f$ and $g$ that satisfy Property
\ref{prop:Incr}, there exists an \lzsette-based compression
algorithm that computes the $(f,g)$-optimal parsing of $S$ in
$O(n(Q(f,n)+ Q(g,n)) + T_{sort}(n,\sigma))$ time and $O(n)$ space in
the worst case.
\end{theorem}

\subsection{On-the-fly generation of $d$-maximal edges}
\label{sub:fsg}

We concentrate only on the computation of the $d$-maximal edges,
because this is the {\em hardest} task. In fact, we know that the
edges outgoing from $v_i$ can be partitioned in no more than
$Q(f,n)$ groups according to the distance from $S[i]$ of the copied
string they represent (proof of Lemma \ref{lem:Qedges}). Let $I_1,
I_2, \ldots, I_{Q(f,n)}$ be the intervals of distances such that all
distances in $I_k$ are encoded with the same number of bits by $f$.
Take now the $d$-maximal edge $(v_i,v_{h_k})$ for the interval
$I_k$. We can infer that substring $S[i:h_k-1]$ is the {\em longest}
substring having a copy at distance within $I_k$ because, by
Definition \ref{def:maximal} and Fact \ref{fact:FS}, any edge
following $(v_i,v_{h_k})$ denotes a longer substring which must lie
in a subsequent interval (by $d$-maximality of $(v_i,v_{h_k})$), and
thus must have longer distance from $S[i]$. Once $d$-maximal edges
are known, the computation of the $\ell$-maximal edges is then easy
because it suffices to further decompose the edges between
successive $d$-maximal edges, say between $(v_i,v_{{h_{k-1}}+1})$
and $(v_i, v_{h_k})$, according to the distinct values assumed by
the encoding function $g$ on the lengths in the range $[h_{k-1},
\ldots, h_k-1]$. This takes $O(1)$ time per $\ell$-maximal edge,
because it needs some algebraic calculations and can then infer the
corresponding copied substring as a prefix of $S[i:h_k-1]$.

So, let us concentrate on the computation of $d$-maximal edges
outgoing from vertex $v_i$. This is based on two key
ideas. The first idea aims at optimizing the space usage
by achieving the optimal $O(n)$ working-space bound. It consists of
proceeding in $Q(f,n)$ passes, one per interval $I_k$ of possible
$d$-costs for the edges in $\prun(S)$. During the $k$th pass, we
logically partition the vertices of $\prun(S)$ in blocks of $|I_k|$
contiguous vertices, say $v_{i_k}, v_{i_k+1}, \ldots,
v_{i_k+|I_k|-1}$, and compute all $d$-maximal edges which spread
from that block and have distance within $I_k$ (thus the same
$d$-cost $c(I_k)$). These edges are kept in memory until they are
used by {\tt Optimal-Parser}, and discarded as soon as the first
vertex of the next block, i.e. $v_{i_k+|I_k|}$, needs to be
processed. The next block of $|I_k|$ vertices is then fetched and
the process repeats. Actually, all passes are executed in {\em
parallel} to guarantee that all $d$-maximal edges of $v_i$ are
available when processing it. There are $n/|I_k|$ distinct blocks,
each vertex belongs to exactly one block at each pass, and all of
its $d$-maximal edges are considered in some pass (because they have
$d$-cost in some $I_k$). The space is
$\sum_{k=1}^{Q(f,n)} |I_k|= O(n)$ because we keep one $d$-maximal
edge per vertex at any pass.

The second key idea aims at computing the $d$-maximal edges for that
block of $|I_k|$ contiguous vertices in $O(|I_k|)$ time and space.
This is what we address in the rest of this paper, because its
solution will allow us to state that the time complexity of {\tt
FSG} is $\sum_{k=1}^{Q(f,n)} \sum _{i=1} ^{n/|I_k|} O(|I_k|) = O(n
\: Q(f,n))$, namely $O(1)$ amortized time per $d$-maximal edge.
Combining this fact with the above observation on the computation of
the $\ell$-maximal edges, we get Theorem \ref{teo:general}.

\begin{figure}[t]
\begin{center}
\includegraphics[scale=0.6]{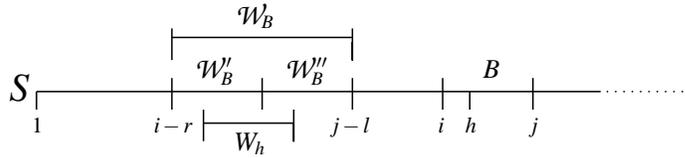}
\end{center}
\caption{The figure shows the interval $B=[i,j]$ with $j=i+|I_k|-1$,
and the window $\W_B$ and its two halves $\W_B'$ and $\W_B''$.
\label{fig:Window}}
\end{figure}

So, let us assume that the alphabet size $\sigma \leq n$, and
consider the $k$th pass of {\tt FSG} in which we assume that
$I_k=[l,r]$. Recall that all distances in $I_k$ can be $f$-encoded
in the same number of, say, $c(I_k)$ bits. Let $B=[i,i+|I_k|-1]$ be
the block of (indices of) vertices for which we wish to compute {\em
on-the-fly} the $d$-maximal edges of cost $c(I_k)$. This means that
the $d$-maximal edge from vertex $v_h$, $h\in B$, represents a
phrase that starts at $S[h]$ and has a copy whose starting position
is in the \textit{window} $W_{h}=[h-r,h-l]$. Thus the distance of
that copy can be $f$-encoded in $c(I_k)$ bits, and so we will say
that the edge has $d$-cost $c(I_k)$. Since this computation must be
done for all vertices in $B$, it is useful to consider $\W_B = W_{i}
\cup W_{i+|I_k|-1}$ which merges the first and last window and thus
spans all positions that can be the (copy-)reference of any
$d$-maximal edge outgoing from $B$. Note that $|\W_B|= 2|I_k|$ (see
Figure \ref{fig:Window}).

The following fact is crucial to efficiently compute all these
$d$-maximal edges via proper indexing data structures:

\begin{fact}
\label{fact:dMax} If there exists a $d$-maximal edge outgoing from
$v_h$ having $d$-cost $c(I_k)$, then this edge can be found by
determining a position $s \in W_h$ whose suffix $S_s$ shares the
{\em maximum} longest common prefix (shortly, \lcp) with $S_h$.
\end{fact}
\begin{Nproof}
Among all positions $s$ in $W_h$ take one whose suffix $S_{s}$ shares
the maximum \lcp\ with $S_h$, and let $q$ be the length of this
\lcp. Of course, there may exist many such positions, we take just one of them. Then the edge
$(v_h,v_{h+q+1})$ has $d$-cost $c(I_k)$ and is $d$-maximal. In fact
any other position $s' \in W_h$ induces the edge $(v_h,v_{h+q'+1})$,
where $q' < q$ is the length of the \lcp\ shared between $S_{s'}$
and $S_h$. This edge cannot be $d$-maximal because its $d$-cost is
still $c(I_k)$ but its length is shorter.
\end{Nproof}

In the rest of the paper we will call the position $s$ of Fact
\ref{fact:dMax} {\em maximal position} for vertex $v_h$. The maximal
position for $v_h$ does exist only if $v_h$ has a $d$-maximal edge
of cost $c(I_k)$. Therefore we design an algorithm which computes
the maximal positions of every vertex $v_h$ in $B$ whenever they do
exist, otherwise it will assign an {\em arbitrary} position to
$v_h$. The net result will be that we will be generating a {\em
supergraph} of $\prun(S)$ which is still guaranteed to have the size
stated in Lemma \ref{lem:Qedges} and can be created efficiently in
$O(|I_k|)$ time and space, as we required above.

Fact \ref{fact:dMax} has related the computation of maximal
positions for the vertices in $B$ to \lcp-computations between
suffixes in $B$ and suffixes in $\W_B$. Therefore it is natural to
resort some indexing data structure, like the compact trie $\T_B$,
built over the suffixes of $S$ which start in the range of positions
$B\cup \W_B$. Trie $\T_B$ takes $O(|B| + |\W_B|) = O(|I_k|)$ space,
and that is within our required space bounds. We then notice that
the maximal position $s$ for a vertex $v_h$ in $B$ having $d$-cost
$c(I_k)$ can be computed by {\em finding the leaf of $\T_B$ which is
labeled with an index $s \in W_h$ and has the deepest lowest common
ancestor (shortly, \lca) with the leaf labeled $h$.} We need to
answer this query in $O(1)$ amortized time per vertex $v_h$, since
we aim at achieving an $O(|I_k|)$ time complexity over all vertices
in $B$. However, this is tricky. In fact this is {\em not} the
classic \lca-query because we do not know $s$, which is actually the
position we are searching for. Furthermore, since the leaf $s$ is
the closest one to $h$ in $\T_B$ among the leaves with index in
$W_h$, one could think to use proper predecessor/successor queries
on a suitable dynamic set of suffixes in $W_h$. Unfortunately, this
would take $\omega(1)$ time because of well-known lower bounds
\cite{beame-fich}. Therefore, answering this query in constant
(amortized) time per vertex requires to devise and deploy
proper structural properties of the trie $\T_B$ and the problem at
hand. This is what we do in our algorithm, whose underlying
intuition follows.

Let $u$ be the \lca\ of the leaves $h$ and $s$ in $\T_B$. For
simplicity, we assume that interval $W_h$ strictly precedes $B$ and
that $s$ is the unique maximal position for $v_h$ (our algorithm
deals with these cases too, see the proof of Lemma
\ref{lem:correctness}). We observe that $h$ must be the smallest
index that lies in $B$ and labels a leaf descending from $u$ in
$\T_B$. In fact assume, by contradiction, that a smaller index $h' <
h$ does exist. By definition $h' \in B$ and thus $v_h$ would not
have a $d$-maximal edge of $d$-cost $c(I_k)$ because it could copy
from the closer $h'$ a possibly longer phrase, instead of copying
from the farther set of positions in $W_h$. This observation implies
that we have to search only for one maximal position per node $u$ of
$\T_B$, and this position refers to the vertex $v_{a(u)}$ whose
index $a(u)$ is the smallest one that lies in $B$ and labels a leaf
descending from $u$. Computing $a$-values clearly takes $O(|\T_B|) =
O(|I_k|)$ time and space.

Now we need to compute the maximal position for $v_{a(u)}$, for each
node $u \in \T_B$. We cannot traverse the subtree of $u$ searching
for the maximal position of $v_{a(u)}$, because this would take
quadratic time complexity. Conversely, we define $\W_B'$ and
$\W_B''$ to be the first and the second half of $\W_B$,
respectively, and observe that any window $\W_h$ has its left
extreme in $\W_B'$ and its right extreme in $\W_B''$. (See Figure
\ref{fig:Window} for an illustrative example.) Therefore the window
$W_{a(u)}$ containing the maximal position $s$ for $v_{a(u)}$
overlaps both $\W_B'$ and $\W_B''$. So if $s$ does exist, then $s$
belongs to either $\W_B'$ or to $\W_B''$, and leaf $s$ descends from
$u$. Therefore, the maximum (resp. minimum) among the elements in
$\W_B'$ (resp. $\W_B''$) that label leaves descending from $u$ must
belong to $W_{a(u)}$. This suggests to compute $\mm(u)$ and $\MM(u)$
as the rightmost position in $\W_B'$ and the leftmost position in
$\W_B''$ that label leaves descending from $u$, respectively. These
values can be computed in $O(|I_k|)$ time by a post-order visit of
$\T_B$.

We are now ready to compute $\maxpos[h]$ as the maximal position for
$v_h$, if it exists, or otherwise set $\maxpos[h]$ arbitrarily. We
initially set all $\maxpos$'s entries to \nil; then we visit $\T_B$ in
post-order and perform, at each node $u$, the following checks
whenever $\maxpos[a(u)]= \nil$:

\begin{smallitemize}
\item If $\mm(u)\in W_{a(u)}$, set $\maxpos[a(u)] = \mm(u)$.
\item If $\MM(u)\in W_{a(u)}$, set $\maxpos[a(u)] = \MM(u)$.
\end{smallitemize}

At the end of the visit, if $\maxpos[a(u)]$ is still $\nil$ we set
$\maxpos[a(u)] = a(\p(u))$ whenever $a(u)  \neq a(\p(u))$. This last
check is needed (proof below) to manage the case in which $S[a(u)]$
can copy the phrase starting at its position from position
$a(\p(u))$ and, additionally, we have that $B$ overlaps $\W_B$
(which may occur depending on $f$). Since $\T_B$ has size
$O(|I_k|)$, the overall algorithm requires $O(|I_k|)$ time and space
in the worst case, as required. The following lemma proves its
correctness:

\begin{lemma} \label{lem:correctness}
For each position $h \in B$, if there exists a $d$-maximal edge
outgoing from $v_h$ and having $d$-cost $c(I_k)$, then $\maxpos[h]$
is equal to its maximal position.
\end{lemma}
\begin{Nproof}
Recall that $B=[i,i+|I_k|-1]$ and consider the longest path $\pi=u_1
u_2 \ldots u_z$ in $\T_B$ that starts from the leaf $u_1$ labeled
with $h\in B$ and goes upward until the traversed nodes satisfy
$a(u_j)=h$, here $j=1,\ldots, z$. By definition of $a$-value, we
know that all leaves descending from $u_z$ and occurring in $B$ are
labeled with an index which is larger than $h$. Clearly, $a(\p(u_z))
< h$ (if any). There are two cases for the final value stored in
$\maxpos[h]$.

Suppose that $\maxpos[h] \in W_h$. We want to prove that
$\maxpos[h]$ is the index of the leaf which has the deepest \lca\
with $h$ among all the other leaves in $W_h$. Let $u_x \in \pi$ be
the node in which the value of $\maxpos[h]$ is assigned (it is
$a(u_x)=h$). Assume that there exists at least another index in
$W_h$ whose leaf has a deeper \lca\ with leaf $h$. This \lca\ must
lie on $u_1 \ldots u_{x-1}$, say $u_l$. Since $W_h$ is an interval
having its left extreme in $\W'_B$ and its right extreme in
$\W''_B$, the value $\MM(u_l)$ or $\mm(u_l)$ must lie in $W_h$ and
thus the algorithm has set $\maxpos[h]$ to one of these positions,
because of the post-order visit of $\T_B$. Therefore $\maxpos[h]$
must be the index of the leaf having the deepest \lca\ with $h$, and
thus by Fact \ref{fact:dMax} is its maximal position (if any).

Now suppose that $\maxpos[h] \notin W_h$ and, thus, it cannot be a
maximal position for $v_h$. We have to prove that it does not exist
a $d$-maximal edge outgoing from the vertex $v_h$ with cost
$c(I_k)$. Let $S_s$ be the suffix in $W_{h}$ having the maximum
\lcp\ with $S_h$, and let $l$ be the \lcp-length. Values $\mm(u_i)$
and $\MM(u_i)$ do not belong to $W_h$, for any node $u_i \in \pi$
(with $a(u_i)=h$), otherwise $\maxpos[h]$ would have been assigned
with an index in $W_h$ (contradicting the hypothesis). The value of
$\maxpos[h]$ remains $\nil$ up to node $u_z$. This implies that no
suffix descending from $u_z$ starts in $W_h$ and, in particular,
$S_s$ does not descend from $u_z$. Therefore, the \lca\ between
leaves $h$ and $s$ is a node in the path from $\p(u_z)$ to root, and
the $\lcp(S_{a(\p(u_z))}, S_h) \geq \lcp(S_s, S_h) = l$. Since
$a(\p(u_z)) < a(u_z)$ and belongs to $B$, it is nearer to $h$ than
any other position in $W_h$, and shares a longer prefix with $S_h$.
So we found longer edge from $v_h$ with smaller $d$-cost. This
implies that $v_h$ has no $d$-maximal edge of cost $c(I_k)$ in
$\prun(S)$.
\end{Nproof}

We are left with the problem of building $\T_B$ in $O(|I_k|)$ time
and space, thus a time complexity which is {\em independent} of the
length of the indexed suffixes and the alphabet size. In Appendix~A
we show how to achieve this result by deploying the fact that the
above algorithm does not make any assumption on the ordering of
$\T_B$, because it just computes (sort of) \lca-queries on its
structure. This is the last step to prove Theorem \ref{teo:general}.

\section{Conclusions}
\label{sec:conclusion}

Our parsing scheme can be extended to variants of \lzsette\ which
deploy parsers that refer to a {\em bounded} compression-window (the
typical scenario of \gzip\ and its derivatives \cite{Salomon04}). In
this case, \lzsette\ selects the next phrase by looking only at the
most recent $w$ input symbols. Since $w$ is usually a constant
chosen as a power of $2$ of few Kbs \cite{Salomon04}, the running
time of our algorithm becomes $O(n\: Q(g,n))$, since $Q(f,w)$ is a
constant. We notice that the remaining term could further be refined
by considering the length $\ell$ of the {\em longest repeated
substring} in $S$, and state the time complexity as $O(n \:
Q(g,\ell))$. If $S$ is generated by an ergodic source
\cite{Spankowski} and $g$ is taken to be the classic Elias' code,
then $Q(g, \ell) = O(\log \log n)$ so that the complexity of our
algorithm results $O(n \log \log n)$ time and $O(n)$ space for this
class of strings.

We finally notice that, although we have mainly dealt with the
\lzsette-dictionary, the techniques presented in this paper could be
extended to design efficient bit-optimal compressors for other
on-line dictionary construction schemes, like \lzotto. Intuitively,
we can show that Theorem \ref{teo:pruning} still holds for any
suffix- or prefix-complete dictionary under the hypothesis that the
codewords assigned to each suffix or prefix of a dictionary phrase
$w$ are {\em shorter than} the codewords assigned to $w$ itself. In
this case the notion of edge maximality (Definition
\ref{def:maximal}) can be generalized by calling an edge $(v_i,v_j)$
maximal iff all longer edges, say $(v_i,v_{j'})$ with $j < j'$, have
not larger cost, namely $w_{i,{j'}} \leq w_{i,j}$. In this case, we
can provide an efficient bit-optimal parser for the
\lzotto-dictionary (details in the full paper).

The main open question is to extend our results to {\em statistical}
encoding functions like Huffman or Arithmetic coders applied on the
integral range $[n]$ \cite{Witten:1999:MGC}. They do not necessarily
satisfy Property \ref{prop:Incr} because it might be the case that
$|f(x)| > |f(y)|$, whenever the integer $y$ occurs more frequently
than the integer $x$ in the parsing of $S$. We argue that it is not
trivial to design a bit-optimal compressor for these
encoding-functions because their codeword lengths change as it
changes the set of distances and lengths used in the parsing
process.

Practically, we would like to implement our optimal parser motivated
by the encouraging experimental results of \cite{Mignosi07}, which
have improved the standard \lzsette\ by a heuristic that {\em tries}
to optimize the encoding cost of {\em just} phrases' lengths.

\bibliographystyle{plain}

\bibliography{optimal}

\begin{thebibliography}{10}

\bibitem{beame-fich}
P.~Beame and F.~E. Fich.
\newblock Optimal bounds for the predecessor problem.
\newblock In {\em Proceedings of the 31st ACM Symposium on Theory of
  Computing}, pages 295--304, 1999.

\bibitem{heur3}
J.~BÃ©kÃ©si, G.~Galambos, U.~Pferschy, and G.J. Woeginger.
\newblock Greedy algorithms for on-line data compression.
\newblock {\em Journal of Algorithms}, 25(2):274--289, 1997.

\bibitem{Vitanyi03}
R.~Cilibrasi and P.~M.~B. Vit{\'a}nyi.
\newblock Clustering by compression.
\newblock {\em CoRR}, cs.CV/0312044, 2003.

\bibitem{heur4}
M.~Cohn and R.~Khazan.
\newblock Parsing with prefix and suffix dictionaries.
\newblock In {\em Data Compression Conference}, pages 180--189, 1996.

\bibitem{CLR}
T.~H. Cormen, C.~E. Leiserson, R.~L. Rivest, and C.~Stein.
\newblock {\em Introduction to Algorithms, Second Edition}.
\newblock The MIT Press and McGraw-Hill Book Company, 2001.

\bibitem{elias}
P.~Elias.
\newblock Universal codeword sets and representations of the integers.
\newblock {\em IEEE Transactions on Information Theory}, 21(2):194--203, 1975.

\bibitem{univ}
P.~Fenwick.
\newblock Universal codes.
\newblock In {\em Lossless Compression Handbook}, pages 55--78. {Academic
  Press}, 2003.

\bibitem{Mignosi07}
A.~Langiu G.~Della~Penna, F.~Mignosi and A.~Ulisse.
\newblock On dictionary-symbolwise data compression.
\newblock In {\em {\tt http://www.di.univaq.it/\%7mignosi/ulicompressor.php}},
  2006.

\bibitem{Raita89}
J.~Katajainen and T.~Raita.
\newblock An approximation algorithm for space-optimal encoding of a text.
\newblock {\em Computer Journal}, 32(3):228--237, 1989.

\bibitem{Raita92}
J.~Katajainen and T.~Raita.
\newblock An analysis of the longest match and the greedy heuristics in text
  encoding.
\newblock {\em Journal of the ACM}, 39(2):281--294, 1992.

\bibitem{klein}
S.~T. Klein.
\newblock Efficient optimal recompression.
\newblock {\em Comput. J.}, 40(2/3):117--126, 1997.

\bibitem{koma00}
R.~Kosaraju and G.~Manzini.
\newblock Compression of low entropy strings with {Lempel--Ziv} algorithms.
\newblock {\em SIAM Journal on Computing}, 29(3):893--911, 1999.

\bibitem{FP99}
Y.~Matias and S.C. \c{S}ahinalp.
\newblock On the optimality of parsing in dynamic dictionary based data
  compression.
\newblock In {\em Procs of the 10th ACM-SIAM Symposium on Discrete Algorithms},
  pages 943--944, 1999.

\bibitem{NM07}
G.~Navarro and V.~M{\"a}kinen.
\newblock Compressed full-text indexes.
\newblock {\em ACM Comput. Surv.}, 39(1), 2007.

\bibitem{SAtax}
S.~J. Puglisi, W.~F. Smyth, and A.~H. Turpin.
\newblock A taxonomy of suffix array construction algorithms.
\newblock {\em ACM Computing Surveys}, 39(2), 2007.

\bibitem{RS:02}
N.~Rajpoot and C.~Sahinalp.
\newblock {\em Handbook of Lossless Data Compression}, chapter Dictionary-based
  data compression, pages 153--167.
\newblock Academic Press, 2002.

\bibitem{Salomon04}
D.~Salomon.
\newblock {\em Data Compression: the Complete Reference, 3rd Edition}.
\newblock Springer Verlag, 2004.

\bibitem{Schuegraf}
E.~J. Schuegraf and H.~S. Heaps.
\newblock A comparison of algorithms for data base compression by use of
  fragments as language elements.
\newblock {\em Information Storage and Retrieval}, 10(9-10):309–319, 1974.

\bibitem{heur1}
M.~E.~Gonzalez Smith and J.~A. Storer.
\newblock Parallel algorithms for data compression.
\newblock {\em Journal of the ACM}, 32(2):344--373, 1985.

\bibitem{Spankowski}
W.~Szpankowski.
\newblock Asymptotic properties of data compression and suffix trees.
\newblock {\em IEEE Transactions on Information Theory}, 39(5):1647--1659,
  1993.

\bibitem{VitterK96}
J.~S. Vitter and P.~Krishnan.
\newblock Optimal prefetching via data compression.
\newblock {\em Journal of the ACM}, 43(5):771--793, 1996.

\bibitem{Witten:1999:MGC}
I.~H. Witten, A.~Moffat, and T.~C. Bell.
\newblock {\em Managing Gigabytes: Compressing and Indexing Documents and
  Images}.
\newblock Morgan Kaufmann Publishers, Los Altos, CA 94022, USA, second edition,
  1999.

\bibitem{lz77}
J.~Ziv and A.~Lempel.
\newblock A universal algorithm for sequential data compression.
\newblock {\em IEEE Transaction on Information Theory}, 23:337--343, 1977.

\bibitem{lz78}
J.~Ziv and A.~Lempel.
\newblock Compression of individual sequences via variable-rate coding.
\newblock {\em IEEE Transactions on Information Theory}, 24(5):530--536, 1978.

\end{thebibliography}

\newpage \appendix
\section*{APPENDIX A. Building $\T_B$ optimally}
\label{sub:trie_build}

We are left with the problem of building $\T_B$ in $O(|I_k|)$ time
and space, thus a time complexity which is {\em independent} of the
length of the indexed suffixes and the alphabet size. We show how to
achieve this result by deploying the crucial fact that the algorithm
of Section \ref{sub:fsg} to compute the $d$-maximal edges does not
make any assumption on the ordering of edges in $\T_B$, because it
just computes (sort of) \lca-queries on its structure. This is the
last step needed to complete the proof of Theorem \ref{teo:general},
and we give its algorithmic details here.

At preprocessing time we build the suffix array of the whole string
$S$ and a data structure that answers constant-time \lcp-queries
between pair of suffixes (see e.g. \cite{SAtax}). These data
structures can be built in $O(n)$ time and space, when $\sigma =
O(n)$. For larger alphabets, we need to add $T_{sort}(n,\sigma)$
time, which takes into account the cost of sorting the symbols of
$S$ and re-mapping them to $[n]$ (see Theorem \ref{teo:general}).

Let us first assume that $B$ and $\W_B$ are contiguous and form the
range $[i, i+3|I_k|-1]$. If we had the sorted sequence of suffixes
starting in $S[i, i+3|I_k|-1]$, we could easily build $\T_B$ in
$O(|I_k|)$ time and space by deploying the above \lcp-data
structure. Unfortunately, it is unclear how to obtain from the
suffix array of the whole $S$, the sorted sub-sequence of suffixes
{\em starting} in the range $[i, i+3|I_k|-1]$ by taking
$O(|B|+|\W_B|)=O(|I_k|)$ time (notice that these suffixes have
length $\Theta(n-i)$). We cannot perform a sequence of
predecessor/successor queries because they would take $\omega(1)$
time each \cite{beame-fich}. Conversely, we resort the key
observation above that $\T_B$ does not need to be ordered, and thus
devise a solution which builds an {\em unordered} $\T_B$ in
$O(|I_k|)$ time and space, passing through the construction of the
suffix array of a {\em transformed} string. The transformation is
simple. We first map the distinct symbols of $S[i, i+3|I_k|-1]$ to
the first $O(|I_k|)$ integers. This mapping {\em does not need to}
reflect their lexicographic order, and thus can be computed in
$O(|I_k|)$ time by a simple scan of those symbols and the use of a
table $M$ of size $\sigma < n$. Then, we define $S^{M}$ as the
string $S$ which has been transformed by re-mapping some of the
symbols according to table $M$ (namely, those occurring in $S[i,
i+3|I_k|-1]$). We can prove that

\begin{lemma}\label{lem:sorting}
Let $S_i, \ldots, S_j$ be a contiguous sequence of suffixes in $S$.
The remapped suffixes $S^M_i \ldots S^M_{j}$ can be
lexicographically sorted in $O(j-i+1)$ time.
\end{lemma}
\begin{Nproof}
Consider the string of pairs $w = \langle S^M[i],b_i\rangle\ldots
\langle S^M[j],b_j\rangle\$$, where $b_h$ is $1$ if $S^M_{h+1} >
S^M_{j+1}$, $-1$ if $S^M_{h+1} < S^M_{j+1}$, or $0$ if $h = j$. The
ordering of the pairs is defined component-wise, and we assume that
$\$$ is a special ``pair'' larger than any other pair in $w$. For
any pair of indices $p,q \in [1 \ldots j-i]$, it is  $S^M_{p+i} >
S^M_{q+i}$ iff $w_{p} > w_{q}$. In fact, suppose that $w_{p} >
w_{q}$ and set $r = \lcp(w_{p},w_{q})$. We have that $w[p+r] =
\langle S^M[p+i+r],b_{p+i+r}\rangle > \langle
S^M[q+i+r],b_{q+i+r}\rangle = w[q+i+r]$. Hence $S^M_{p+i+r} >
S^M_{q+i+r}$, by definition of the $b$'s. Therefore $S^M_{p+i} >
S^M_{q+i}$, since their first $r$ symbols are equal. This implies
that sorting $S^M_i,\ldots,S^M_j$ reduces to computing the suffix
array of $w$, and this takes $O(|w|)$ time given that the alphabet
size is $O(|w|)$ \cite{SAtax}. Clearly, $w$ can be constructed in
that time bound because comparing $S^M_z$ with $S^M_{j+1}$ takes
$O(1)$ time via an \lcp-query on $S$ (using the proper data
structure above) and a check at their first mismatch.
\end{Nproof}

Lemma \ref{lem:sorting} allows us to generate the compact trie of
$S^M_i,\ldots, S^M_{i+3|I_k|-1}$, which is equal to the (unordered)
compacted trie of $S_i,\ldots,S_{i+3|I_k|-1}$ after replacing every
{\tt ID} assigned by $M$ with its original symbol in $S$. We finally
notice that if $B$ and $\W_B$ are not contiguous (as instead we
assumed above), we can use a similar strategy to sort separately the
suffixes in $B$ and the suffixes in $\W_B$, and then merge these two
sequences together by deploying the \lcp-data structure mentioned at
the beginning of this section.

\end{document}